\documentclass[aps,showpacs,nofootinbib,twocolumn]{revtex4}
\usepackage[dvips]{graphicx}
\usepackage{amssymb}
\newcommand{\ped}[1]{\ensuremath{_{\rm #1}}}
\newcommand{\apex}[1]{\ensuremath{^{\rm #1}}}

\begin{document}

\title{A study of carbon substitutions in MgB$_2$ within the two-band Eliashberg theory}

\author{G.A. Ummarino}\email{E-mail:giovanni.ummarino@infm.polito.it}
\author{D. Daghero}
\author{R.S. Gonnelli}

\affiliation{ Dipartimento di Fisica, Politecnico di Torino, Corso
Duca degli Abruzzi 24, 10129 Torino, Italy} \affiliation{INFM-
LAMIA, Corso Perrone 24, 16152 Genova, Italy}

\author{A.H. Moudden}
\affiliation{Laboratoire L\'{e}on Brillouin, CEA-CNRS, CE Saclay,
91191 Gif-sur-Yvette, France}

\begin{abstract}
We study the effects of C substitutions in MgB$_2$ within the
two-band model in the Eliashberg formulation. We use as input the
$B-B$ stretching-mode frequency and the partial densities of states
$N\ped{N}\apex{\sigma}(E\ped{F})$ and
$N\ped{N}\apex{\pi}(E\ped{F})$, recently calculated for
Mg(B$\ped{1-x}$C$\ped{x}$)$_2$ at various $x$ values from
first-principles density functional methods. We then take the
prefactor in the Coulomb pseudopotential matrix, $\mu$, and the
interband scattering parameter, $\Gamma\apex{\sigma \pi}$, as the
only adjustable parameters. The dependence on the C content of
$T\ped{c}$ and of the gaps ($\Delta\ped{\sigma}$ and
$\Delta\ped{\pi}$) recently measured in
Mg(B$\ped{1-x}$C$\ped{x}$)$_2$ single crystals indicate an almost
linear decrease of $\mu$ on increasing $x$, with an increase in
\emph{interband} scattering that makes the gaps merge at $x=0.132$.
In polycrystals, instead, where the gap merging is not observed, no
interband scattering is required to fit the experimental data.
\end{abstract}


\maketitle

In spite of its simple structure, the intermetallic compound MgB$_2$
-- discovered to be superconducting at about 40~K in 2001
\cite{Nagamatsu} -- soon revealed a number of surprising features
that could not be explained within a picture of conventional
superconductivity. Bandstructure calculations \cite{bandstructure}
showed that the energy bands of MgB$_2$ can be grouped into two
sets: the quasi-2D $\sigma$ bands, and the 3D $\pi$ bands,
originating from the superposition of in-plane and out-of-plane
boron orbitals, respectively. As a matter of fact, most of the
physical properties of this superconductor have found a clear and
relatively simple explanation within an \emph{effective} two-band
model \cite{Liu,Brinkman,Choi} in which the two bands interact via a
phonon-mediated \emph{interband coupling}. The result is that
superconductivity develops in both bands at the same $T\ped{c}$, but
with energy gaps of different amplitude, $\Delta\ped{\sigma}$ and
$\Delta\ped{\pi}$, and thus different values of the gap ratio
$2\Delta/k\ped{B}T\ped{c}$. The success of the two-band model in
describing the features of MgB$_2$ naturally opens the question
whether it can predict (or at least explain a posteriori) the
effects of induced disorder, irradiation and, over all, chemical
substitutions on the physical properties of the compound. As far as
substitutions are concerned, the experimental test of theoretical
predictions has been delayed or even prevented by the technical
difficulties in obtaining good-quality samples of partially
substituted MgB$_2$ \cite{substitutions}. Recently, point-contact
measurements of the gap amplitudes as a function of the C content
have been reported in state-of-the-art
Mg(B$\ped{1-x}$C$\ped{x}$)$\ped{2}$ polycrystals \cite{Holanova} and
single crystals \cite{nostroC}. The availability of these results
(that for some aspects contrast with each other) gives a good
opportunity to test the two-band model. In this paper we will show
that both the experimental data concerning $T\ped{c}$ and the gaps
as a function of $x$ can be well explained within the two-band model
in the Eliashberg formulation. We will use as input the frequencies
of the B-B stretching mode (which is strongly coupled to the holes
in the $\sigma$ band) and the partial densities of states at the
Fermi level, $N\ped{N}^{\sigma}(E\ped{F})$ and
$N\ped{N}^{\pi}(E\ped{F})$, calculated from first-principle density
functional methods adopting the viewpoint of ordered supercells
\cite{Moudden_paper} instead of the virtual-crystal approximation.
Then, we will show that the experimental $x$ dependence of
$T\ped{c}$ and of the gaps $\Delta\ped{\sigma}$ and
$\Delta\ped{\pi}$ can be very well reproduced by admitting a
reasonable $x$ dependence of the prefactor in the Coulomb
pseudopotential matrix \cite{Brinkman,Golubov} and, in the case of
single crystals, an increase in the \emph{interband scattering}
$\Gamma\apex{\sigma \pi}$ on increasing the C content.

Let us start from the generalization of the Eliashberg theory
\cite{Eliashberg,Marsiglio} for systems with two bands
\cite{Kresin}, that has already been used with success to study
the MgB$_{2}$ system
\cite{Brinkman,Choi,Golubov,John1,Mazin0,John3}. To obtain the
gaps and the critical temperature within the $s$-wave, two-band
Eliashberg model one has to solve four coupled integral equations
for the gaps $\Delta_{i}(i\omega_{n})$ and the renormalization
functions $Z_{i}(i\omega_{n})$, where $i$ is a band index and
$\omega\ped{n}$ are the Matsubara frequencies. We included in the
equations (explicitly reported elsewhere \cite{John1}) the
non-magnetic impurity scattering rates in the Born approximation,
$\Gamma\apex{ij}$.

The solution of the Eliashberg equations requires as input: i) the
four (but only three independent\cite{Kresin}) electron-phonon
spectral functions $\alpha^{2}_{ij}(\omega)F(\omega)$; ii) the
four (but only three independent\cite{Kresin}) elements of the
Coulomb pseudopotential matrix $\mu^{*}(\omega\ped{c})$; iii) the
two (but only one independent \cite{Kresin}) effective impurity
scattering rates $\Gamma\apex{ij}$.
%
%%%%%%%%%%%%%%%%%%%%%%%%%%%%%%%%%%%%%%%%%%%%%%%%%%%%%%%%%%%%%%%%%%%%
\begin{figure}[t]
\begin{center}
\includegraphics[keepaspectratio,width=0.7\columnwidth]{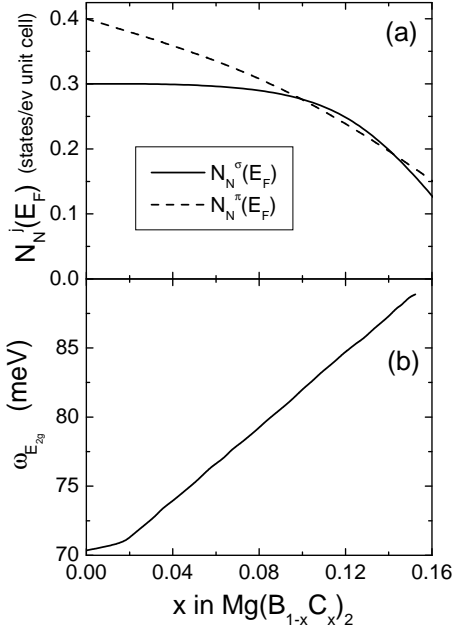}
\end{center}
\vspace{-5mm} \caption{(a) Calculated density of states at the
Fermi energy, $N\ped{N}\apex{\sigma}(E\ped{F})$ (solid line) and
$N\ped{N}\apex{\pi}(E\ped{F})$ (dashed line) as a function of $x$
(From Ref. \onlinecite{Moudden_paper}). (b) Calculated frequency
of the B-B bond stretching mode (the $E\ped{2g}$ mode in pure
MgB$_2$) as a function of $x$ (from
Ref.~\onlinecite{Moudden_paper}). } \label{fig:Moudden}
\end{figure}
%%%%%%%%%%%%%%%%%%%%%%%%%%%%%%%%%%%%%%%%%%%%%%%%%%%%%%%%%%%%%%%%%%%%%
%
None of these parameters or functions has been calculated for
C-substituted MgB$_2$, and in many cases their determination is a
very difficult task, at least at the present moment. Hence, we are
left with a problem with too many degrees of freedom. However, we
will now show how some reasonable approximations allow reducing
the number of adjustable parameters to 2, with no significant loss
of generality.

%
%%%%%%%%%%%        discussione a2F(omega)        %%%%%%%%
%

\begin{figure}[t]
\begin{center}
\includegraphics[keepaspectratio, width=\columnwidth]{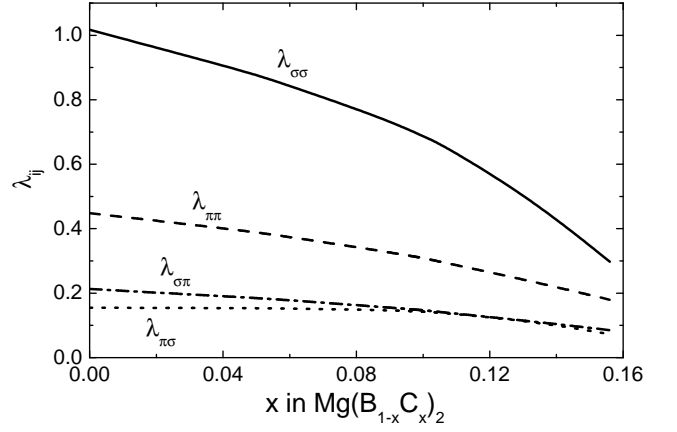}
\end{center}
\vspace{-5mm} \caption{Electron-phonon coupling constants
$\lambda_{ij}$ calculated as a function of $x$ according to
eqs.~\ref{eq:lambda_ss} and \ref{eq:lambda_ij}. }
\label{fig:lambda}
 \end{figure}

Let's start with the four spectral functions
$\alpha^{2}_{ij}(\omega)F(\omega)$, that were calculated for pure
MgB$_{2}$ in ref. \onlinecite{Golubov}. For simplicity, we will
assume that the shape of the $\alpha^{2}_{ij}F(\omega,x)$ functions
does not change with $x$, and we will only rescale them with the
electron-phonon coupling constants $\lambda_{ij}$:
\begin{equation}
\alpha^{2}_{ij}F(\omega,x)=
\frac{\lambda_{ij}(x)}{\lambda_{ij}(x=0)}
\alpha^{2}_{ij}F(\omega,x=0)\label{eq:a2F}
\end{equation}
Neglecting the effect of C substitution on the shape of the e-ph
spectral functions is not a dramatic simplification, since we
showed in a previous paper \cite{John1} that the details of
$\alpha^{2}F(\omega)$ do not significantly affect the resulting
$T_c$. To determine the scaling factor in eq.~\ref{eq:a2F}, let us
remind that, from the definition of electron-phonon coupling
constant \cite{Grimvall}:
\begin{equation}
\lambda=\frac{N\ped{N}(E\ped{F})<I^{2}>}{M \Omega^{2}_{0}}
\label{eq:lambda}
\end{equation}
where $M$ is the ion mass, $\Omega_{0}$ is a frequency
representative of the phonon spectrum, $N\ped{N}(E\ped{F})$ is the
density of states at the Fermi level and $<I^{2}>$ is the average
matrix element of the electron-ion interaction \cite{Grimvall}. In
our case, $M$ is the boron mass \cite{Liu} and does not depend on
$x$. As a first approximation, and as we did in the case of Al
substitution \cite{John1}, we will assume that also the average
matrix element of the electron-ion interaction $<I^{2}>$ is
constant, because it is basically determined by the deformation
potential which is almost independent of $x$ \cite{Profeta}. The
partial densities of states at the Fermi level,
$N\ped{N}\apex{\sigma}(E\ped{F})$ and
$N\ped{N}\apex{\pi}(E\ped{F})$, have been recently calculated from
first principles by using a supercell approach \cite{Moudden_paper}
for different values of the C content $x$, and are reported in
Fig.\ref{fig:Moudden}(a). The frequency $\Omega_{0}$ can be
identified with the frequency of the B-B bond-stretching phonon mode
(the $E\ped{2g}$ mode), that has been recently calculated as a
function of $x$ from first principles \cite{Moudden_paper}, and is
reported in Fig.\ref{fig:Moudden}(b). Since this mode couples
strongly with the holes on top of the $\sigma$ band, from
eq.~\ref{eq:lambda} we will have for $\lambda\ped{\sigma \sigma}$
(which gives the most important contribution to superconductivity in
our system):
\begin{equation}
\lambda\ped{\sigma\sigma}(x)
=\frac{N\ped{N}\apex{\sigma}(E\ped{F},x)\omega^{2}\ped{E\ped{2g}}(x=0)}
{N\ped{N}\apex{\sigma}(E\ped{F},x=0) \omega^{2}\ped{E\ped{2g}}(x)}
\lambda\ped{\sigma\sigma}(x=0).\label{eq:lambda_ss}
\end{equation}
In this way, we assume that the change in the frequency of the
$E\ped{2g}$ phonon affects the coupling constant, while we neglect
its influence on the shape of the electron-phonon spectral
function. For the other coupling constants, we will instead assume
for simplicity
\begin{equation}
\hspace{-1mm}\forall (i,j) \neq (\sigma,\sigma) \hspace{5mm}
\lambda\ped{ij}(x)
=\frac{N\ped{N}\apex{j}(E\ped{F},x)}{N\ped{N}\apex{j}(E\ped{F},x=0)}
\lambda\ped{ij}(x=0)\label{eq:lambda_ij}
\end{equation}
with $\lambda_{\sigma\sigma}(x\!\!=\!\!0)$=1.017,
$\lambda_{\pi\pi}(x\!\!=\!\!0)$=0.448,
$\lambda_{\sigma\pi}(x\!\!=\!\!0)$=0.213 and
$\lambda_{\pi\sigma}(x\!\!=\!\!0)$=0.155 \cite{Golubov,Brinkman}.
Fig.~\ref{fig:lambda} shows the calculated electron-phonon coupling
constants $\lambda_{ij}$ as a function of $x$.
%
%%%%%%%%%%%%%%%%%%%%%%%%%%%%%%%%%%%%%%%%%%%%%%%%%%%%%%%%%%%%%%%%%%%%%
%%%%%%%%%%%%%%%%%%%%%%%%%%%%%%%%%%%%%%%%%%%%%%%%%%%%%%%%%%%%%%%%%%%%%

%%%%%%%%%%%%%%    discussione mu   %%%%%%%%%%%%%%
%
As far as the Coulomb pseudopotential is concerned, let us start
from its expression in pure MgB$_2$
\cite{Brinkman,Golubov,Dolgov}:
\begin{eqnarray}
\hspace{-5mm}\mu^{*}(x)\!\!&\!=\!&\!\! \left| \begin{array}{cc}%
\mu^{*}\ped{\sigma \sigma} & \mu^{*}\ped{\sigma \pi}\\
\mu^{*}\ped{\pi \sigma} & \mu^{*}\ped{\pi \pi}
\end{array} \right| =  \nonumber \\
\!\!&\!=\!&\!\! \mu(\omega_{c},x)N\ped{N}\apex{tot}(E\ped{F},x)
\left| \begin{array}{cc}%
\frac{2.23}{N\ped{N}\apex{\sigma}(E\ped{F},x)} &
\frac{1}{N\ped{N}\apex{\sigma}(E\ped{F},x)}\\ & \\
\frac{1}{N\ped{N}\apex{\pi}(E\ped{F},x)} &
\frac{2.48}{N\ped{N}\apex{\pi}(E\ped{F},x)}
\end{array} \right| \label{eq:mu}
\end{eqnarray}
where $\mu(\omega\ped{c},x)$ is a free parameter and
$N\ped{N}\apex{tot}(E\ped{F},x)$ is the total normal density of
states at the Fermi level. The numbers 2.23 and 2.48 in the Coulomb
matrix have been calculated for pure MgB$_{2}$ but, as a first
approximation, we will suppose them not to depend on $x$. In this
way, the elements of the Coulomb pseudopotential matrix,
$\mu^{*}\ped{ij}$, depend on $x$ only through the densities of
states at the Fermi level and through the common prefactor
$\mu(\omega_{c},x)$, that could also take into account the effects
of disorder.
%
%%%%%%%%%%%%%%%%     discussione Gamma  %%%%%%%%%%%%%%%%

As far as the scattering rates are concerned, let us remind that,
due to Anderson's theorem, \emph{intraband} scattering does not
affect neither $T\ped{c}$ nor the gaps \cite{Mazin}, so we will
disregard both $\Gamma\apex{\sigma \sigma}$ and $\Gamma\apex{\pi
\pi}$. The remaining interband scattering parameters are related
to each other since \cite{Kresin}
\begin{equation}
\frac{\lambda\ped{ij}(x)}{\lambda\ped{ji}(x)}=\frac{\mu^{*}%
\ped{ij}(x)}{\mu^{*}\ped{ji}(x)}=%
\frac{\Gamma\apex{ij}(x)}{\Gamma\apex{ji}(x)}=\frac{N\ped{N}%
\apex{j}(E\ped{F},x)}{N\ped{N}\apex{i}(E\ped{F},x)}
\label{eq:Shulga}
\end{equation}
and thus we will always refer only to $\Gamma\apex{\sigma \pi}$.
Finally, we can fix the cut-off energy (e.g., $\omega_{c}=700$ meV)
so as to reduce the number of adjustable parameters to two: the
prefactor in the Coulomb pseudopotential, $\mu(\omega_{c})$ (that we
will call simply $\mu$ from now on) and the interband scattering
parameter $\Gamma\apex{\sigma\pi}$.

%
%%%%%%%%%%%%%%%%%%%%%%%%%%%%%%%%%%%%%%%%%%%%%%%%%%%%%%%%%%%%%%%%%%%%%
\begin{figure}[t]
\begin{center}
\includegraphics[keepaspectratio, width=\columnwidth]{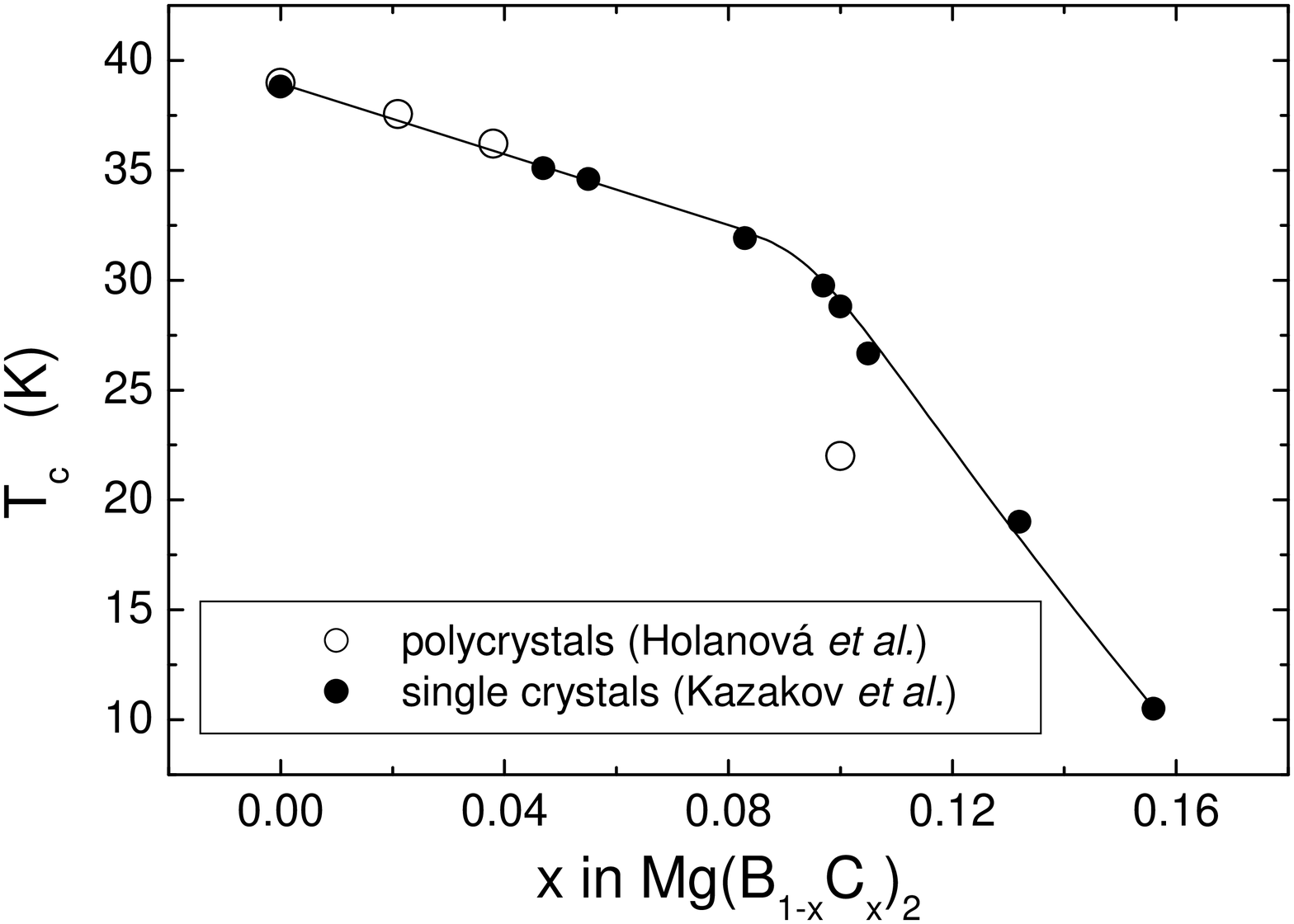}
\end{center}
\vspace{-5mm} \caption{The experimental $T_{c}$ measured in
Mg(B$\ped{1-x}$C$\ped{x}$)$_2$ single crystals \cite{Kazakov} (solid
circles) and polycrystals \cite{Holanova} (open circles) as a
function of $x$. The line is only a guide to the eye.}
\label{fig:Tc}
\end{figure}
\begin{figure}[t]
\begin{center}
\includegraphics[keepaspectratio, width=\columnwidth]{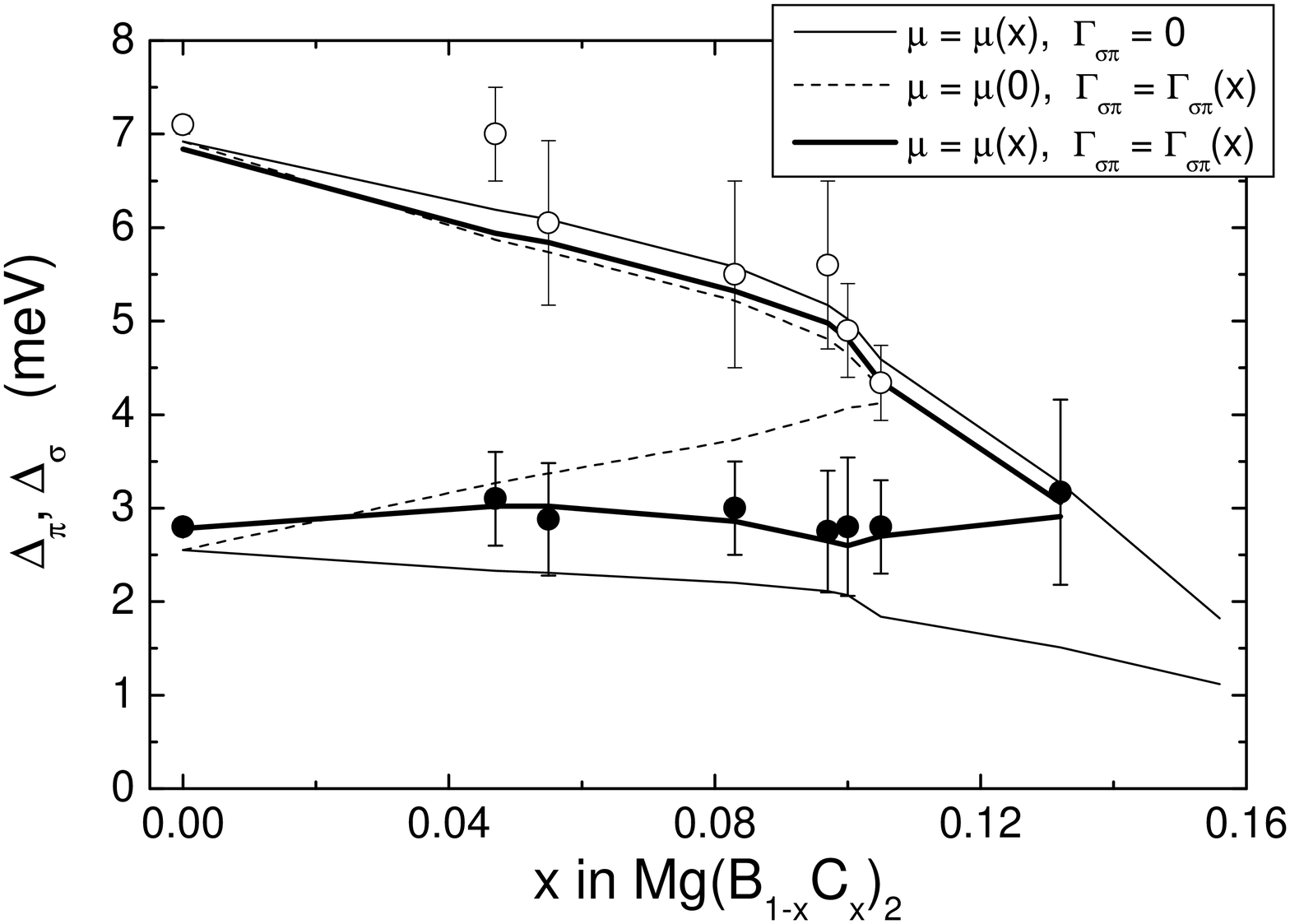}
\end{center}
\vspace{-5mm} \caption{Symbols: experimental values of
$\Delta_{\sigma}$ (open circles) and $\Delta_{\pi}$ (solid
circles) measured at $T=4.2$~K by PCS in single crystals (from
Ref.~\onlinecite{nostroC}). Lines: $\Delta_{i}\,(i\omega_{n=0})$
calculated for the $\sigma$ and $\pi$ bands at $T=T_{c}/4$ by
solving the imaginary-axis Eliashberg equations in the following
cases: i) thin solid line: $\Gamma\apex{\sigma \pi}$=0, and $\mu$
varies with $x$ as shown in Fig.~\ref{fig:gamma_mu}(a), open
circles; ii) dashed line: $\mu(x)=\mu(0)$ and $\Gamma\apex{\sigma
\pi}$ varies with $x$ as shown in Fig.~\ref{fig:gamma_mu}(b), open
squares; iii) thick solid line: both $\mu$ and $\Gamma\apex{\sigma
\pi}$ vary with $x$ as shown in Fig.~\ref{fig:gamma_mu}(a) and
(b), respectively (solid symbols).}\label{fig:gap_crystals}
\end{figure}
\begin{figure}[t]
\includegraphics[keepaspectratio, width=\columnwidth]{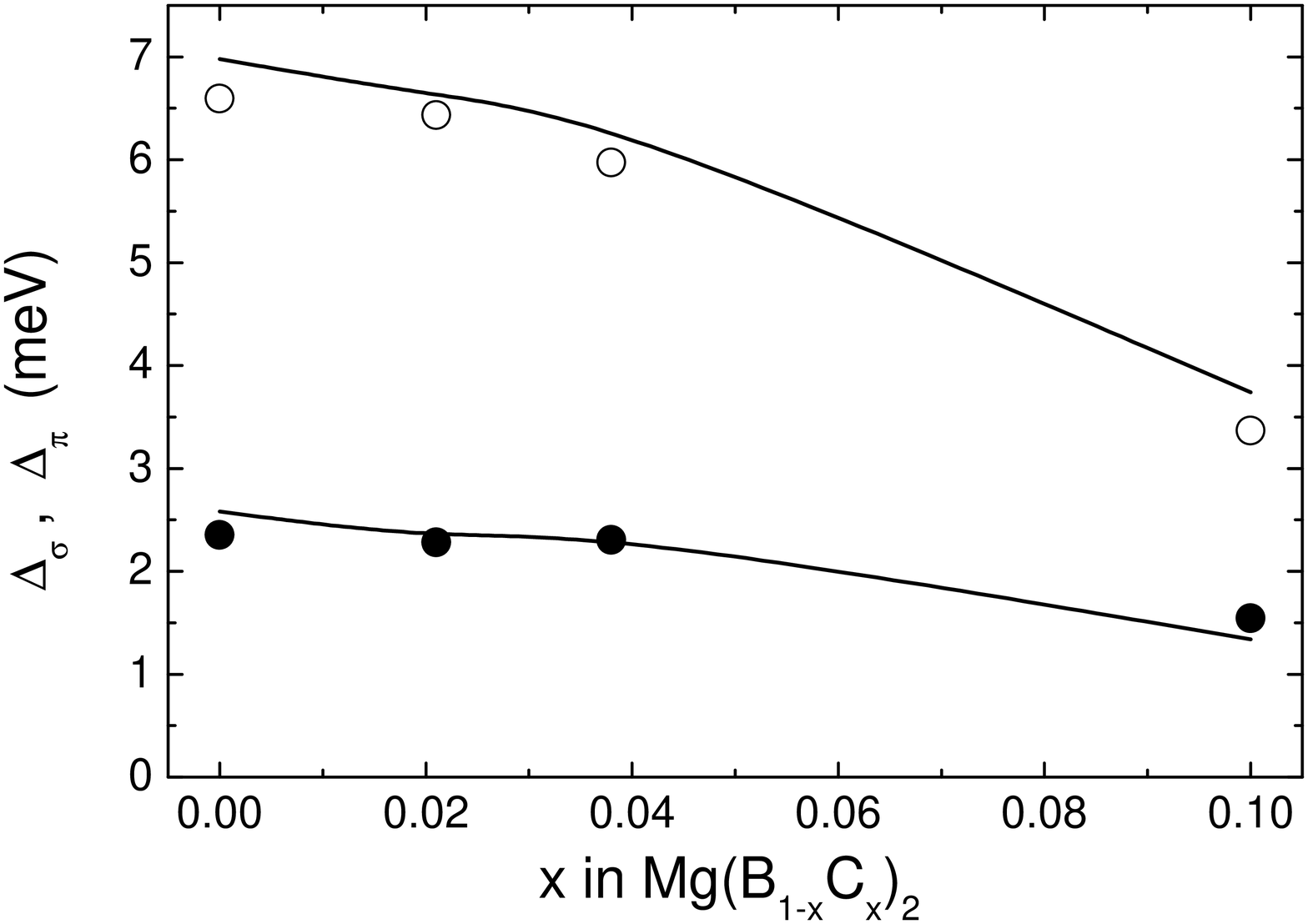}
\caption{Experimental values of $\Delta_{\sigma}$ (open circles)
and $\Delta_{\pi}$ (solid circles) measured  by PCS at $T=4.2$~K
(from Ref. \onlinecite{Holanova}) compared to the values of
$\Delta_{i}(i\omega_{n=0})$ of the $\sigma$ and $\pi$ bands
(lines) calculated by solving the imaginary-axis Eliashberg
equations at $T=T_{c}/4$, when all the physical parameters vary
with $x$ apart from the interband scattering which is kept equal
to zero.} \label{fig:gap_poly}
\end{figure}

%%%%%%%%%%%%%%%%%%%%%%%%%%%%%%%%%%%%%%%%%%%%%%%%%%%%%%%%%%%%%%%%%%%%%
%
%%%%%%%%%%%%%%     fine discussione del modello       %%%%%%%%%%%%%%%%%%%
%
As already pointed out, the aim of the present work is to show that
the experimental dependence of $T\ped{c}$ and of the gaps,
$\Delta\ped{\sigma}$ and $\Delta\ped{\pi}$, on the C content in
Mg(B$\ped{1-x}$C$\ped{x}$)$_2$ can be explained within the two-band
Eliashberg theory. The experimental $T\ped{c}(x)$ curves measured in
single crystals \cite{Kazakov} and polycrystals \cite{Holanova} are
reported in Fig.\ref{fig:Tc}. The corresponding $x$ dependencies of
the gaps measured by point-contact spectroscopy (PCS) are reported
in Fig.~\ref{fig:gap_crystals} and Fig.~\ref{fig:gap_poly},
respectively (symbols). In single crystals
(Fig.~\ref{fig:gap_crystals}), the two gaps approach each other on
increasing $x$, until at $x=0.132$ they become experimentally
indistinguishable. This means that, at this doping content, their
amplitudes are equal to each other within the experimental
uncertainty. In polycrystals, instead, the two gaps remain clearly
distinct up to $x=0.10$, where $\Delta\ped{\pi}$ is much smaller
than in single crystals with the same C content (see
Fig.~\ref{fig:gap_poly}).

Let us focus for the time being on single crystals. The
$T\ped{c}(x)$ curve (solid circles in Fig.~\ref{fig:Tc}) can be
exactly reproduced by adjusting only one of the two free parameters
of the model, or both of them at the same time (but, in this case,
the choice of their values is not univocal unless one adds another
constraint).

\begin{figure}[t]
\begin{center}
\includegraphics[keepaspectratio, width=0.8\columnwidth]{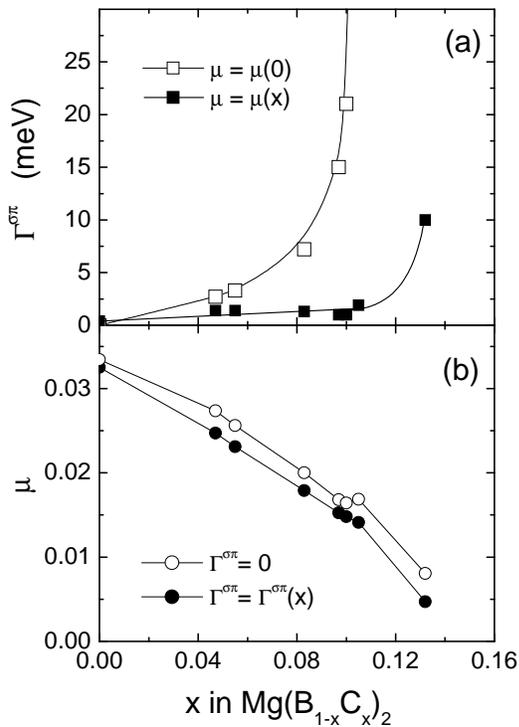}
\end{center}
\vspace{-5mm}\caption{(a) Open squares: the $x$ dependence of
$\Gamma\apex{\sigma \pi}$ necessary to reproduce the $T\ped{c}$ of
single crystals, if $\mu(x) = \mu(0)$. Solid squares: the
$\Gamma\apex{\sigma \pi}(x)$ curve that allows fitting both
$T\ped{c}$ and the gaps when also $\mu$ is varied with $x$. Lines
are only guides to the eye.  (b) Open circles: the $x$ dependence of
the prefactor in the Coulomb pseudopotential, $\mu$, that gives the
$T\ped{c}(x)$ curve measured in single crystals (solid symbols in
Fig.\ref{fig:Tc}), when $\Gamma\apex{\sigma \pi}=0$. Solid circles:
the $\mu(x)$ curve that allows fitting both $T\ped{c}$ and the gaps
($\Delta\ped{\sigma}$ and $\Delta\ped{\pi}$) in single crystals,
when also $\Gamma\apex{\sigma \pi}$ is varied.} \label{fig:gamma_mu}
\end{figure}

For example, one can keep $\mu$ equal to its value in pure MgB$_2$
(i.e., $\mu(x)=\mu(0)$), and view the substituted compound as a
``disordered'' version of MgB$_2$, as if the only effect of C
substitution was an increase in interband scattering. This implies
neglecting also the phonon hardening and the electron-doping
effects (that actually play a leading role in determining the
observed $T\ped{c}(x)$ curve \cite{Kortus_nuovo}) so that the
$T\ped{c}(x)$ curve is reproduced by only varying
$\Gamma\apex{\sigma \pi}$. The resulting trend of the interband
scattering rate is shown in Fig.\ref{fig:gamma_mu}(a) (open
squares). Notice that with this approach one cannot obtain
critical temperatures lower than $T\ped{c}=25.8$~K, that
corresponds to the isotropic ``dirty'' limit in which the two gaps
merge into one of amplitude $\Delta=4.1$~meV \cite{Liu}. This is
clearly seen in the $x$ dependence of the gaps calculated with
these values of $\Gamma\apex{\sigma \pi}$, which is reported in
Fig.~\ref{fig:gap_crystals} as a dashed line. In spite of a rather
good agreement between experimental and theoretical values of
$\Delta\ped{\sigma}$, the model predicts an increase in
$\Delta\ped{\pi}$ which is not observed, and the merging of the
two gaps at a much lower C content with respect to the actual one.

%
%%%%%%%%%%%%%%%%%%%%%%%%%%%%%%%%%%%%%%%%%%%%%%%%%%%%%%%%%%%%%%%%%%%%%
%
%%%%%%%%%%%%%%%%%%%%%%%%%%%%%%%%%%%%%%%%%%%%%%%%%%%%%%%%%%%%%%%%%%%%%
%
The opposite case consists in taking into account all the effects of
substitutions (i.e. phonon hardening and electron doping), with no
increase of interband scattering. In this case, one can keep
$\Gamma\apex{\sigma \pi}=0$, and vary $\mu$ with $x$ so as to
reproduce the experimental $T\ped{c}(x)$ curve. With the resulting
values of $\mu(x)$, shown in Fig.\ref{fig:gamma_mu}(b) as open
circles, one obtains the $x$ dependence of the gaps indicated in
Fig.~\ref{fig:gap_crystals} as thin solid lines. It is clear that
the $\mu(x)$ curve that reproduces the experimental $T\ped{c}$ for
any C content gives values of the large gap $\Delta\ped{\sigma}$
that agree rather well with the experimental ones (open circles in
Fig.~\ref{fig:gap_crystals}) but gives rise to a decrease in the
small gap which is not observed experimentally.

%%%%%%%%%%%%%%%%%%%%%%%%%%%%%%%%%%%%%%%%%%%%%%%%%%%%%%%%%%%%%%%%%%%%%
%%%%%%%%%%%%%%%%%%%%%%%%%%%%%%%%%%%%%%%%%%%%%%%%%%%%%%%%%%%%%%%%%%%%%%
%
The analysis of the previous two cases suggests that the
experimental $\Delta\ped{\sigma}(x)$ and $\Delta\ped{\pi}(x)$ curves
could be explained as due to the interplay between a decrease in
$\mu$ (that makes $\Delta\ped{\pi}$ decrease) and an increase in
$\Gamma\apex{\sigma \pi}$ (that instead makes $\Delta\ped{\pi}$
increase). This result has been recently anticipated by an analysis
of the effects of band filling and interband scattering
\cite{Kortus_nuovo}. Hence, we will now try to fit the experimental
$x$ dependence of $T\ped{c}$ and of the gaps $\Delta\ped{\pi}$ and
$\Delta\ped{\sigma}$ by varying \emph{both} $\mu$ and
$\Gamma\apex{\sigma \pi}$. The best-fitting curves for the gaps
(actually, the values of $\Delta(i\omega_{n=0})$ at $T=T_{c}/4$, for
the two bands) are reported as thick solid lines in
Fig.~\ref{fig:gap_crystals}. The choice of the parameters is
univocal, and the resulting $x$ dependencies of $\Gamma\apex{\sigma
\pi}$ and $\mu$ are reported as solid symbols in
Fig.~\ref{fig:gamma_mu}(a) and (b), respectively.

As shown in Fig.~\ref{fig:gamma_mu}(a), the interband scattering
remains smaller than 2~meV (which is a value reasonable even for
pure MgB$_2$) up to $x=0.10$ and then increases to make the gaps
approach each other until they become indistinguishable. The point
at $x=0.132$ in Fig.~\ref{fig:gamma_mu}(a) represents the minimum
value of $\Gamma\apex{\sigma \pi}$ that gives gap values differing
less than 0.3 meV (which is approximately the best experimental
resolution of PCS at 4.2 K). Greater values of $\Gamma\apex{\sigma
\pi}$ are allowed as well, since they would give rise to gaps even
closer to each other. Although the point at $x=0.132$ might depend
on the approximations we are using in the present paper, there is no
doubt that $\Gamma\apex{\sigma \pi}$ has to increase to reproduce
the experimental gap values \cite{Kortus_nuovo}. This increase is
thus a general prediction of the two-band Eliashberg theory, but its
origin in C-substituted MgB$_2$ is still debated at the moment.
According to Ref.~\onlinecite{Erwin}, carbon substitutions should
not change the local lattice point symmetry and therefore the
interband scattering should remain very small as in pure MgB$_2$
\cite{Mazin}. However, a $\sigma$-$\pi$ hybridization might also
arise, above $x=0.10$, from the presence of superstructures or even
short-range order in the substituted compound \cite{Moudden_paper}.
It must be said, however, that high-resolution TEM has shown no
superstructures in these single crystals \cite{Kazakov}, even if the
possibility of short-range order is not ruled out. An alternative
explanation is based on the observed increase in flux pinning and in
the normalized resistance on increasing $x$ \cite{Kazakov}. These
effects suggest the existence of microscopic defects in the single
crystals, acting as scattering centers. As indicated by
magnetization data, these defects might be local inhomogeneities in
the C distribution on a length scale comparable to $\xi$, that may
well give rise also to $\sigma-\pi$ scattering.

\begin{figure}[t]
\begin{center}
\includegraphics[keepaspectratio, width=\columnwidth]{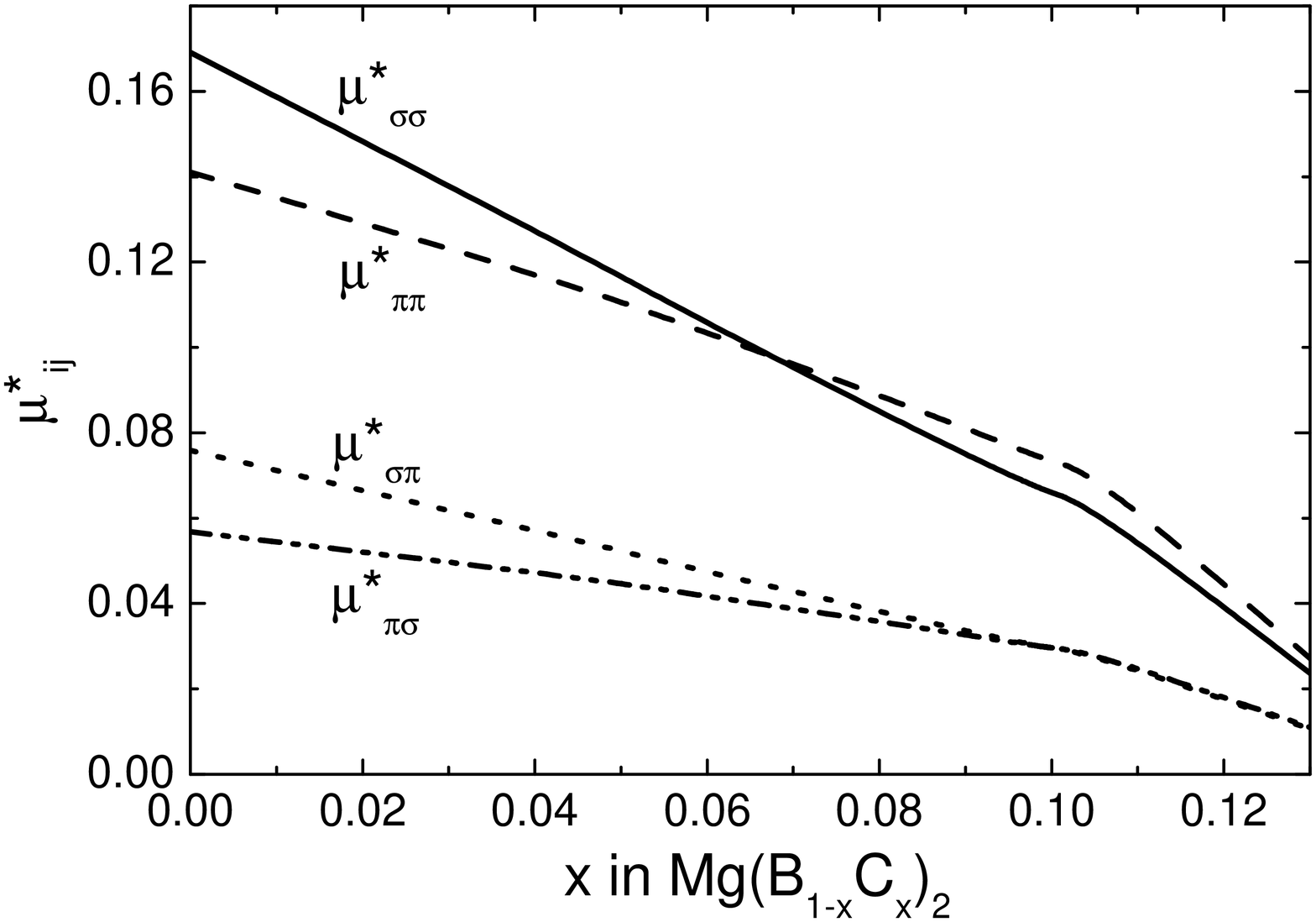}
\end{center}
\caption{The elements of the Coulomb pseudopotential matrix,
$\mu^{*}_{ij}$, calculated from eq.\ref{eq:mu} by using the
densities of states from first-principles calculations and the
prefactor $\mu(x)$ that best fits the experimental data
($T\ped{c}$ and gaps) in Mg(B$\ped{1-x}$C$\ped{x}$)$_2$ single
crystals. } \label{fig:mu_star}
\end{figure}

%%%%%%%%%%%%%%%%%%%%%%%%%%%%%%%%%%%%%%%%%%%%%%%%%%%%%%%%%%%%%%%
% Possible explanation of mu variation
The values of the Coulomb pseudopotential prefactor, $\mu$, that
allow reproducing both the $T\ped{c}$ and the gap amplitudes, are
reported in Fig.~\ref{fig:gamma_mu}(b) (solid circles) as a function
of $x$. The resulting $\mu(x)$ curve is almost linear up to
$x=0.10$, where a change in slope reflects the analogous feature of
the experimental $T\ped{c}$ (see Fig.~\ref{fig:Tc}).
Fig.~\ref{fig:mu_star} reports the values of the components of the
Coulomb pseudopotential matrix, $\mu^{*}\ped{ij}$, calculated from
eq.~\ref{eq:mu} by using the densities of states
($N\ped{N}\apex{\sigma}(E\ped{F},x)$ and
$N\ped{N}\apex{\pi}(E\ped{F},x)$) from density-functional methods,
and the values of $\mu(x)$ that allow best-fitting the experimental
gaps (solid symbols in Fig.~\ref{fig:gamma_mu}(b)). It is clear that
all the components of the $\mu^{*}$ matrix monotonically decrease on
increasing the C content. The large decrease (by a factor of two) of
$\mu$ or, similarly, of $\mu^{*}\ped{\sigma \sigma}$ between $x$=0
and $x$=0.1, suggests large changes in the electronic screening,
that seem to be incompatible with the much smaller changes in the
partial densities of states (Fig.\ref{fig:Moudden}). Giving an
explanation of this puzzle within the two-band model is a very
difficult task. However, a tentative and qualitative explanation can
be given in the much simpler single-band case. Let us therefore
consider the $\sigma$-band quantities alone. Let $\mu^{*}\equiv
\mu^{*}\ped{\sigma \sigma}$ be the
renormalized Coulomb pseudopotential, given by $\mu^{*}=\mu%
[ 1+ \mu \ln ( E\ped{F}/\omega\ped{log})]^{-1}$. Starting from the
value $\mu^{*}(x\!\!=\!\!0)\simeq 0.17 $ (see
Fig.\ref{fig:mu_star}) and using $E\ped{F}$=500~meV
\cite{Pietronero} and $\omega\ped{log}$=$\omega\ped{E2g}$, the
value of the bare Coulomb pseudopotential $\mu=0.26$ is obtained.
From the Morel-Anderson definition \cite{Morel} of $\mu$, i.e.:
\begin{equation}
\mu= \frac{1}{2 (\frac{2k_{F}}{k_{S}})^{2}} \cdot \ln
\left[1+\left(\frac{2k_{F}}{k_{S}}\right)^2\right]
\end{equation}
where $k_S$ is the screening wavevector, and using as a first
approximation the free-electron relationship between $k_{F}$ and
$E_{F}$, one gets $k_{S}(x\!\!=\!\!0)$=0.47~\AA$^{-1}$. The same
calculation gives, for $x=0.1$,
$k_{S}(x\!\!=\!\!0.1)$=0.16~\AA$^{-1}$, so that
$[(k_{S}(x\!\!=\!\!0)/k_{S}(x\!\!=\!\!0.1)]^2=8.56$. Since in the
Morel-Anderson model $k_{S}^2 \propto k_{TF}^2$ (where $k_{TF}$ is
the Thomas-Fermi screening wavevector) and $k_{TF}^2$ is
proportional to $N(E_{F})$, this value has to be compared to the
ratio
$N\ped{N}^{\sigma}(E_{F},x\!\!=\!\!0)/N\ped{N}^{\sigma}(E_{F},x\!\!=\!\!0.1)$=1.11.
The comparison confirms that the change in the DOS alone cannot
explain the observed reduction in $\mu^{*}$. However, a large
increase in the residual resistivity is observed on increasing the C
content~\cite{Kazakov}, so that $\rho\ped{0}(x\!\!=\!\!0.1) \approx
5 \rho\ped{0}(x\!\!=\!\!0)$. This suggests that, for some $x > 0.1$,
a metal-to-insulator (MIT) transition might be expected. In the
hypothesis that at $x=0.1$ the system already lies somewhere between
the Fermi liquid and the critical regime where the MIT occurs, a
generalization of the Morel-Anderson model \cite{Soulen} has to be
used to describe the $x=0.1$ case. Within this scenario, $k_{S}^2
(x\!\!=\!\!0.1) \propto k_{TF}^2 [1+(a/\alpha r)^2]^{-1}$, where
$r=[\rho_{0}(x\!\!=\!\!0.1)-\rho_{c}]/\rho\ped{c}$, $\rho\ped{c}$ is
the critical value of the residual resistivity and $a$, $\alpha$ are
constants defined in Ref.~\onlinecite{Soulen}. Hence one gets
\begin{equation}
\left[\frac{k_{S}(x=0)}{k_{S}(x=0.1)}\right]^2=\frac{N\ped{N}\apex{\sigma}(x=0)}%
{N\ped{N}\apex{\sigma}(x=0.1)}\cdot \left[ 1+ \left(\frac{a}{\alpha
r}\right)^2\right]\label{eq:ks_ratio}
\end{equation}
from which $(a/\alpha r)= 2.78$. The parameter $\alpha r$ expresses
the distance from criticality (i.e. from the MIT) and can be
obtained from \makebox{
$N\ped{N}\apex{\sigma}(E\ped{F},x\!\!=\!\!0.1)=N\ped{N}\apex{\sigma}(E\ped{F},x\!\!=\!\!0)%
[1-\exp(-\alpha r)]$}, that gives $\alpha r=2.3$. According to Ref.
\onlinecite{Soulen}, this value is perfectly compatible with a
strongly disordered Fermi liquid. Finally, the value of the constant
$a$ turns out to be $a=6.4$ that falls in the range of values
allowed in Ref.\onlinecite{Soulen} and is correctly of the order of
the cell parameter. In conclusion, the observed drop of
$\mu^{*}\ped{\sigma \sigma}$ is due to a change in the screening
length that, in turns, can be justified by the transition to a
disordered Fermi liquid on increasing the C content. Incidentally,
this result might further justify the observed increase in interband
scattering $\Gamma\apex{\sigma \pi}$ at high doping levels.

%
%%%%%%%%%%%%%%%%%%%%%%%%%%%%%%%%%%%%%%%%%%%%%%%%%%%%%%%%%%%%%%%%%%%%%
%%%%%%%%%%%%%%%%%%%%%%%%%%%%%%%%%%%%%%%%%%%%%%%%%%%%%%%%%%%%%%%%%%%%%

\begin{figure}[t]
\includegraphics[keepaspectratio, width=\columnwidth]{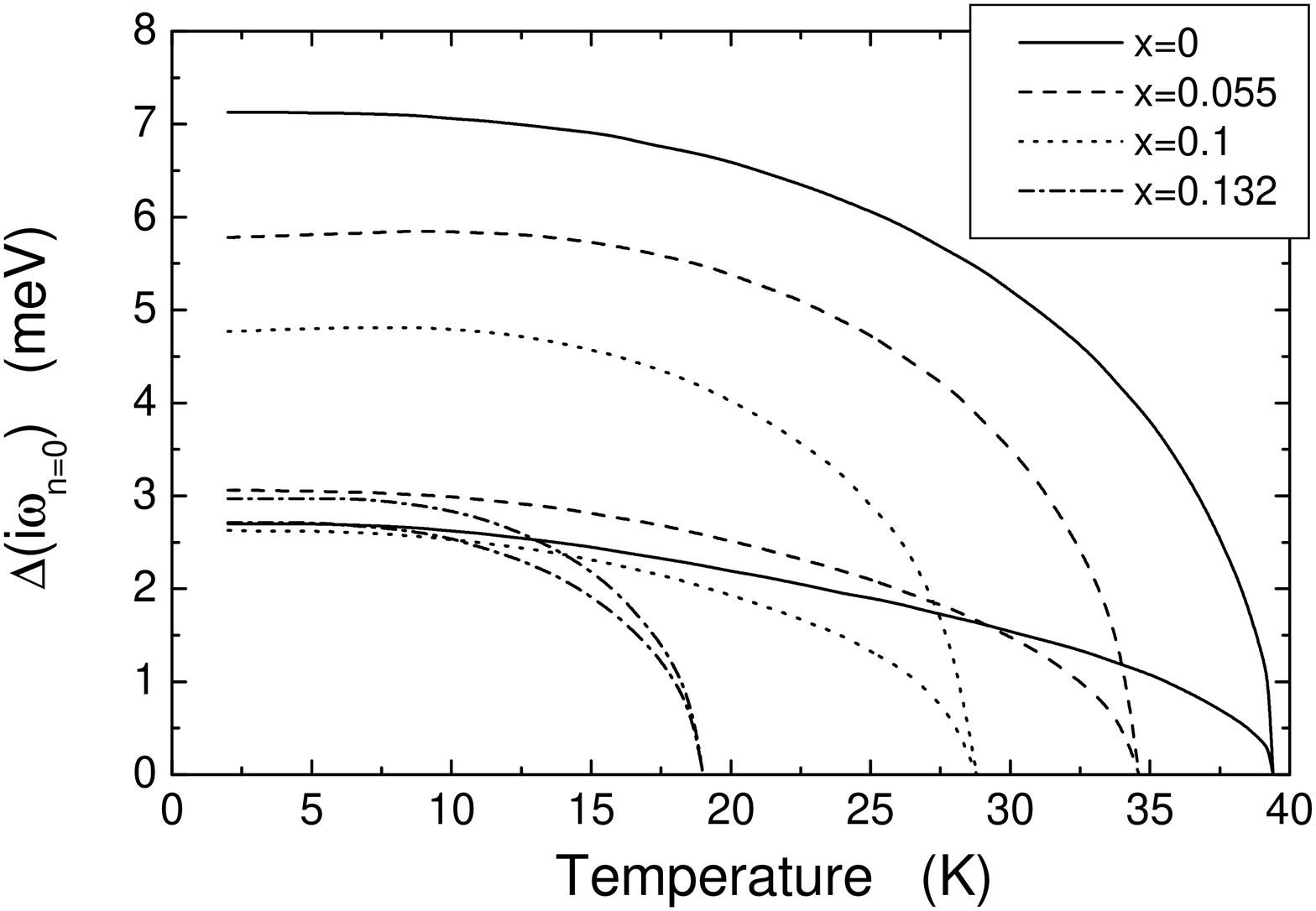}
\caption{The temperature dependence of $\Delta_{\pi}(i\omega_{n=0})$
and $\Delta_{\sigma}(i\omega_{n=0})$ calculated by solving the
Eliashberg equations in four different cases: $x=0$ (solid lines),
$x=0.055$ (dashed lines), $x=0.1$ (dotted lines) and $x=0.132$
(dashed-dotted lines).} \label{fig:gaps(T)}
\end{figure}

At this point, all the parameters entering the two-band model in
Eliashberg formulation have been determined as a function of the C
content, so that in principle any relevant physical property of the
superconducting state of Mg(B$\ped{1-x}$C$\ped{x}$)$_2$ single
crystals can be calculated. For the time being, we can calculate the
temperature dependence of the gaps at different C contents, that can
be easily tested by performing PCS measurements as a function of
temperature.  Fig.~\ref{fig:gaps(T)} reports the calculated values
of $\Delta_{\pi}(i\omega_{n=0})$ and
$\Delta_{\sigma}(i\omega_{n=0})$ as a function of $T$ given by the
solution of the Eliashberg equations in four different cases: $x=0$
(solid lines), $x=0.055$ (dashed lines), $x=0.1$ (dotted lines) and
$x=0.132$ (dash-dotted lines). It is worthwhile to notice that, even
at high C contents, the $\Delta\ped{\pi}(T)$ curve shows a negative
curvature in the whole temperature range as in pure MgB$_2$. This is
due to the fact that, as shown in Fig.~\ref{fig:lambda}, the
interband coupling does not decrease sensibly on increasing $x$ --
otherwise a positive curvature would be observed in
$\Delta\ped{\pi}(T)$ in the proximity of $T\ped{c} $~\cite{Suhl}.
%
%%%%%%%%%%%%%%%%%%%%%%%%%%%%%%%%%%%%%%%%%%%%%%%%%%%%%%%%%%%%%%%%%%%
%%%%%%%%%%%%%%%%%%%%%%%%%%%%%%%%%%%%%%%%%%%%%%%%%%%%%%%%%%%%%%%%%%%
%

Let us now turn our attention to the experimental results obtained
in Mg(B$\ped{1-x}$C$\ped{x}$)$_2$ polycrystals \cite{Holanova}. As
we did in the case of single crystals, we start by trying to
reproduce the experimental $T\ped{c}(x)$ curve (open circles in
Fig.~\ref{fig:Tc}) keeping $\Gamma\apex{\sigma \pi}=0$ and varying
the prefactor in the Coulomb pseudopotential, $\mu$. Once determined
the values of $\mu$ that give exactly the experimental $T\ped{c}$,
we can calculate the gaps $\Delta(i\omega_{n=0})$ at $T=T_{c}/4$ for
the $\sigma$ and $\pi$ bands. The results are reported as a function
of $x$ in  Fig.~\ref{fig:gap_poly} (solid lines).  Surprisingly, the
calculated gaps agree very well with those measured by PCS
(symbols), with no need of interband scattering. This result
indicates that the strong difference between the trend of the gaps
measured in single crystals \cite{nostroC} and polycrystals
\cite{Holanova} is very likely to be due to the different nature of
the samples. Unfortunately, a more detailed discussion would require
a deeper knowledge of the mechanisms that give rise to interband
scattering in C-substituted samples, which is lacking at the present
moment - even though some hypotheses for the increase in
$\Gamma\apex{\sigma \pi}$ in single crystals have been presented
above.
%
%%%%%%%%%%%%%%%%%%%%%%%%%%%%%%%%%%%%%%%%%%%%%%%%%%%%%%%%%%%%%%%%%%%%%
%%%%%%%%%%%%%%%%%%%%%%%%%%%%%%%%%%%%%%%%%%%%%%%%%%%%%%%%%%%%%%%%%%%%
%

In conclusion, we have studied the Mg(B$\ped{1-x}$C$\ped{x}$)$_2$
system within the effective two-band Eliashberg model, that was
already shown to be well suited for the description of unsubstituted
MgB$_2$. In the analysis of the C-substituted system, we have used
as input parameters the frequency of the B-B stretching mode and the
partial densities of state at the Fermi level, calculated as a
function of $x$ by first-principles density-functional methods.
Adopting some reasonable approximations, we have come to a
simplified model with only two adjustable parameters (the prefactor
in the Coulomb pseudopotential and the interband scattering rate),
whose dependence on $x$ has been determined so as to reproduce the
experimental values of $T\ped{c}$ and of the gaps
$\Delta\ped{\sigma}$ and $\Delta\ped{\pi}$.

The success of the model in describing the experimental findings
shows that C-substituted MgB$_2$ is a weak-coupling two-band system
as the parent compound. In details, the results indicate that: i)
the experimental behaviour of $T\ped{c}$ on increasing $x$ is the
results of the decrease in the $\sigma-\sigma$ intraband coupling
(due to the filling of the $\sigma$ bands) and of a decrease in all
the elements of the Coulomb pseudopotential matrix, in particular
$\mu^{*}\ped{\sigma \sigma}$. Without the contribution from
$\mu^{*}$, the $T\ped{c}(x)$ curve would be steeper than
experimentally observed \cite{Kortus_nuovo}; ii) the different trend
of the gaps observed experimentally in single crystals (where the
gaps become indistinguishable at $x=0.132$) and polycrystals (where
there is no tendency to gap merging) only arises from the different
amount of interband scattering in the two cases. The increase in
$\Gamma\apex{\sigma \pi}$ above $x\simeq 0.10$ might arise from
short-range order in the single crystal structures
\cite{Moudden_paper}, or from local inhomogeneities in the C content
on a microscopic scale \cite{Kazakov}.

Finally, these results give an indication of what an ideal
substitution, capable of increasing the $T\ped{c}$ of the MgB$_2$
system, should do, i.e. increase $\lambda\ped{\sigma \sigma}$,
decrease $\mu^{*}\ped{\sigma \sigma}$, and keep the interband
scattering as small as in pure MgB$_2$. According to
eqs.~\ref{eq:lambda_ss} and \ref{eq:mu}, this is possible if
$N\ped{N}\apex{\sigma}(E\ped{F})$ increases and
$N\ped{N}\apex{\pi}(E\ped{F})$ decreases.

Many thanks are due to S. Massidda and A. Bianconi for useful
discussions. This work was done within the Project PRA "UMBRA" of
INFM, the FIRB Project RBAU01FZ2P and the INTAS Project n.01-0617.

\end{document}